\documentclass[aps,prl,twocolumn,groupadress]{revtex4}
\usepackage{graphicx}
\usepackage{amsmath}
\usepackage{amssymb}
\usepackage{epsfig}

\begin{document}

\title{Electric Field Effect Tuning of Electron-Phonon Coupling in Graphene }

\author{Jun Yan,$^{1}$ Yuanbo Zhang,$^{1}$ Philip Kim,$^{1}$ and Aron Pinczuk$^{1,2}$}

\affiliation{$^1$ Department of Physics, Columbia University, New York, NY 10027, USA \\
    $^2$ Department of Applied Physics and Applied Mathematics, Columbia University, New York, NY 10027, USA}

\date{\today}

\begin{abstract}
Gate-modulated low-temperature Raman spectra reveal that the
electric field effect (EFE), pervasive in contemporary electronics,
has marked impacts on long wavelength optical phonons of graphene.
The EFE in this two dimensional honeycomb lattice of carbon atoms
creates large density modulations of carriers with linear dispersion
(known as Dirac fermions). Our EFE Raman spectra display the
interactions of lattice vibrations with these unusual carriers. The
changes of phonon frequency and line-width demonstrate optically the
particle-hole symmetry about the charge-neutral Dirac-point. The
linear dependence of the phonon frequency on the EFE-modulated Fermi
energy is explained as the electron-phonon coupling of mass-less
Dirac fermions.
\end{abstract}


\maketitle

The interaction between electrons and quantized lattice vibrations
in a solid is one of the most fundamental realms of study in
condensed matter physics. In particular, the electron-phonon
interaction in graphene and its derivatives plays an important role
in understanding anomalies of photoemission spectra observed in
graphite \cite{Zhou} and graphene \cite{Rotenberg}, the non-linear
high energy electron transport in carbon nanotubes
\cite{Kane1,Javey,Park,Ando1,Kane2}, as well as phonon structures in
graphite \cite{Ferrari1,Ferrari2} and carbon nanotubes
\cite{Dresselhaus,Ferrari2,Kuzmany}.
\par
\begin{figure}
\centering \epsfig{figure=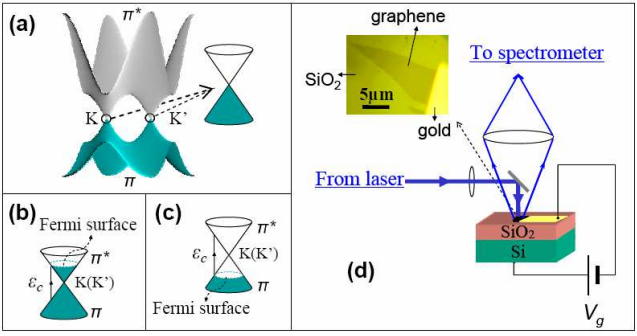, width=0.99\linewidth,clip=}
\caption{(color online). Electronic band structure of graphene (a-c)
and experimental set-up (d). (a) $\pi$ and $\pi$* bands of graphene
in momentum space. K and K' (Dirac points) are the corner points of
the first Brillouin zone. For neutral graphene, the $\pi$ band is
completely filled shown as blue shading in this diagram. Near Dirac
points, the band structure exhibits isotropic linear dispersion
shown as two cones at K and K'. (b) shows the region near Dirac
points in n-type graphene. $\varepsilon_c$ is the onset energy for
vertical electron-hole pair transitions. (c) is like (b) for p-type
graphene. (d) Upper left is an optical image of the sample. }
\label{fig1}
\end{figure}

Traditionally, electron-phonon interactions are investigated through
chemical doping, in which the charge carrier density is varied by
introduction of impurities. The electric field effect (EFE) is an
alternative method for changing the charge carrier density
effectively in low-dimensional systems. The EFE has proven very
successful in graphene, a single atomic sheet of graphite, where
unconventional integer quantum Hall effect \cite{Geim2,Kim} has
revealed physics linked to the uniqueness of the electronic band
structure near the charge neutral Dirac points (Fig. \ref{fig1}(a)).

\par
We measured Raman spectra of optical phonons in graphene where large
densities of free electrons or free holes are modulated by the EFE.
We discovered that the even parity long wavelength optical phonon
(the graphene G band) has marked dependence on gate voltage and the
induced charge density. The dependence of phonon frequency and
line-width on the EFE induced charge density demonstrates that the
intriguing physics of mass-less Dirac fermions with particle-hole
symmetry is encoded in the electron-phonon interaction.

\par
Raman studies of graphite \cite{Thomsen} are at the forefront of
research on carbon based materials. The recent availability of
few-layer and single-layer graphene \cite{Geim1,deHeer}, has
stimulated great interest in Raman scattering in such novel and
exciting systems. For example, dimensional crossover was observed in
Raman spectra of thin graphitic films as a function of multilayer
thickness \cite{Ferrari3,Eklund,Ensslin}. In the work reported here,
Raman spectroscopy emerges as an insightful method to probe the EFE
in a single atomic layer and the phonon dynamics that are associated
with the two dimensional (2D) Dirac fermions.

\par
We focus on the doubly degenerate optical phonon of E$_{2g}$
symmetry at $\sim$1580 cm$^{-1}$, known as the G band. We also
report on the smaller impact of the EFE on the second-order band at
$\sim$2700 cm$^{-1}$, known as the D* band. These two bands are
prominent Raman features in graphene \cite{Ferrari3,Eklund,Ensslin}.
G band is the optical phonon at long wavelengths and D* band is
associated with a two-phonon state in which each phonon has a large
wavevector. Our experiments demonstrate that the G band is markedly
sensitive to coupling with Dirac fermion excitations at small
wavevectors (long wavelengths), while the reduced impact of the EFE
on the D* band offers insights on the coupling to particle-hole
pairs at large wavevectors.

\par
The interaction of the G phonon with small wavevector particle-hole
pairs has a crucial dependence on the onset energy $\varepsilon_c$
for vertical transitions (zero momentum transfer) of an electron
from $\pi$ valence band to a $\pi$* conduction band state as shown
in Figs. \ref{fig1}(b) and \ref{fig1}(c). The $\pi\rightarrow\pi$*
transitions need to satisfy the requirements of the Pauli principle.
For this reason, $\varepsilon_c$ is directly linked to the position
of the Fermi surface as $\varepsilon_c=2|E_F|$, where $|E_F|$ is the
absolute value of the Fermi energy. The response of the G phonon to
the EFE is thus inextricably related to the electronic band
structure in the vicinity of the Dirac point.

\par
Graphene samples were prepared using the mechanical method described
in Ref. \cite{Geim3}. The sample was mounted on degenerately doped
p-type silicon with a 300-nm thick SiO$_{2}$ as gate dielectric. A
gold electrode was thermally evaporated on one end of the graphene
layer to contact it to the electrical ground (Fig. \ref{fig1}(d)). A
gate bias voltage $V_g$ was applied to the silicon substrate to
induce charge carriers in the sample by the EFE. The sample was
mounted in a variable temperature cryostat with optical access where
it was cooled by a continuous flow of cold Helium gas. The Raman
measurements were performed in a backscattering configuration (Fig.
\ref{fig1}(d)) with the 488-nm line of an argon-ion laser as
exciting radiation.  The laser spot size on the sample was $\sim25~
\mu m$ in diameter and the power was kept below 5 mW. The scattered
light was collected into a Spex-1404 0.85-m double grating
spectrometer and spectra were recorded by a liquid nitrogen cooled
CCD camera. The spectral resolution in these measurements is $\sim$2
cm$^{-1}$.

\par
\begin{figure}
\centering \epsfig{figure=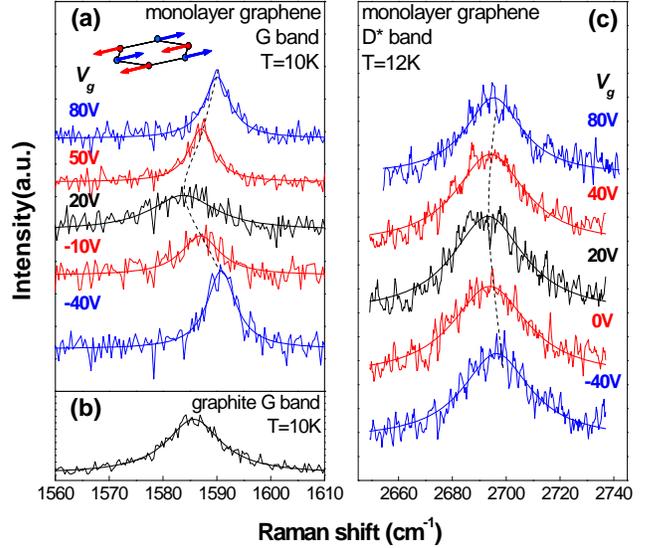, width=0.96\linewidth,clip=}
\caption{(color online). Low temperature Raman spectra and their
Lorentzian fits (smooth overlapping curves). (a) Evolution of the G
band of a graphene monolayer with gate voltage $V_g$. The upper left
inset is a schematic representation of carbon atom motion in the G
band. The dashed line is a guide to the eye. (b) Spectrum of the G
band of a thick ($>$10 nm) graphite layer. (c) Evolution of the D*
two-phonon band of a graphene monolayer with $V_g$. } \label{fig2}
\end{figure}

Typical spectra showing G and D* bands at cryostat cold finger
temperatures $\sim$10 K are displayed in Fig. \ref{fig2}.
Remarkably, both peak position and line-width are found to be
modulated by the gate voltage. The changes induced by the EFE are
much larger for the G mode than for the second-order D* band. We
also show for comparison a G band for thick ($>$10 nm) samples (Fig.
\ref{fig2}(b)) taken under similar experimental conditions with no
gate voltage applied, which serves as a bulk sample spectrum.

\par
The EFE induced charge in the single layer of graphene can be
modulated by gate voltage according to the relation $n =
C_g(V_g-V_{Dirac})/e$, where $C_g$, $e$, and $V_{Dirac}$ are the
gate capacitance, electron charge and the gate voltage corresponding
to the charge-neutral Dirac-point. The changes in graphene Raman
spectra as a function of $V_g$ demonstrate that the EFE modulated
charge density results in significant changes in the frequency and
lifetime of phonons.

\par
\begin{figure}
\centering \epsfig{figure=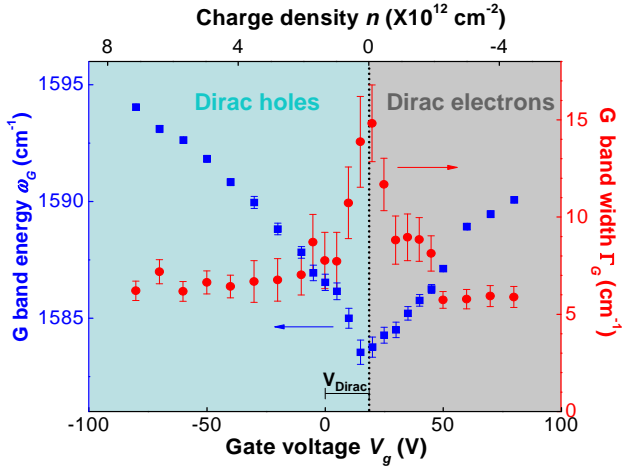, width=0.95\linewidth,clip=}
\caption{(color online). G band energy (squares) and G band width
(circles) extracted from Fig. 2(a). The vertical dotted line is the
approximate position of the charge neutral Dirac point that is
estimated from the symmetry of the data. The upper scale is obtained
with $n=C_g(V_g-V_{Dirac})/e$, where $C_g$=115 aF$/\mu m^2$
\cite{Geim2,Kim}.} \label{fig3}
\end{figure}

Figure \ref{fig3} displays the G-mode frequency $\omega_G$ and width
$\Gamma_G$ as a function of $V_g$.  It is remarkable that
$\omega_G(V_g)$ and $\Gamma_G(V_g)$ both exhibit nearly symmetric
changes relative to the value of $V_g\sim$ 18 V. This symmetry is
naturally linked to the symmetry of the electronic band structure
that occurs at the Dirac point shown in Fig. \ref{fig1}(a), from
which we determine $V_{Dirac} = 18\pm2$ V. In particular,
$\omega_G(V_g)$ has a minimum near the Dirac point, revealing that
increases in charge density of either sign result in stiffening of
the G mode. On the other hand, $\Gamma_G$ sharply decreases as
$|V_g-V_{Dirac}|$ increases, showing that longer phonon lifetimes
are linked to higher particle/hole density. These results offer key
insights into the coupling of Dirac fermions to lattice vibration
modes, as discussed below.

\par
\begin{figure}
\centering \epsfig{figure=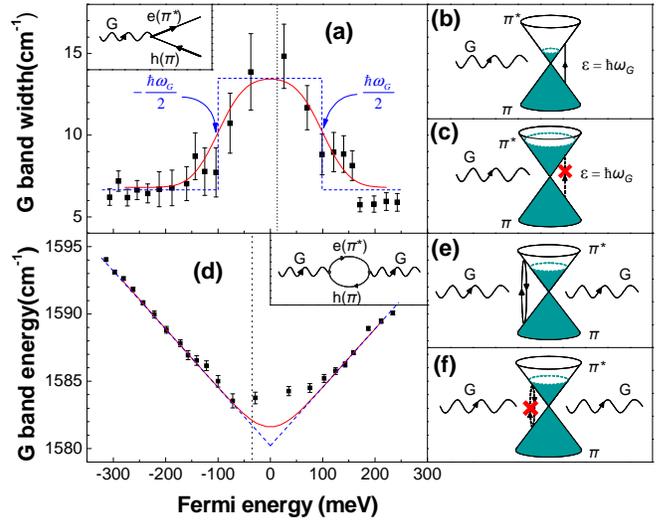, width=0.99\linewidth,clip=}
\caption{(color online). Graphene G band damping (a-c) and energy
renormalization (d-f). In (a) and (d), blue dashed lines and red
solid lines are the fits for ideal and non-uniform graphene
respectively. In all cases, the same electron-phonon coupling
strength ($D=12.6~\textrm{eV/\AA})$ is assumed. The vertical dotted
lines are the determinations of the Dirac point shown in Fig.
\ref{fig3}. The insets are Feynman diagrams for electron-phonon
coupling applicable to the case of the G phonon. (b) represents the
broadening of the G phonon due to decay into particle-hole pairs.
(c) indicates that the G phonon decay into electron-hole pair is
forbidden by the Pauli principle at high EFE induced charge
densities. (e) is for the renormalization of the G phonon energy by
creation of virtual electron-hole pairs. (f) shows that virtual
electron-hole pair transitions with energy ranging from 0 to
$2|E_F|$ are forbidden. Only the diagrams for n-type graphene are
shown in panels (b), (c), (e), and (f). Diagrams for the p-type
cases are similar.} \label{fig4}
\end{figure}

We first consider $\Gamma_G$  shown in Fig. \ref{fig4}(a). Its
marked reduction (from 15 to 6.5 cm$^{-1}$) at higher carrier
densities can be understood as a Landau damping of phonons in which
the mode decays into particle-hole pairs. Examples of such processes
are shown in the inset to Fig. \ref{fig4}(a) and in Fig.
\ref{fig4}(b). Phonon decay processes are real transitions that
conserve energy and momentum. The small wavevector G phonon can only
decay into particle-hole pairs represented by vertical transitions
that have vanishingly small wavevector transfer (Fig.
\ref{fig4}(b)). When charge carriers are induced by EFE in ideal
graphene at low temperatures, the Fermi level $E_F$ changes as
$E_F(n)=-sgn(n)\hbar v_F\sqrt{(\pi|n|)}$ , where $v_F=10^6$m/sec
\cite{Geim2,Kim} is the Fermi velocity and $n$ is charge density. To
satisfy the Pauli principle, Landau damping is allowed only when
$|E_F|<\hbar\omega_G/2$ (Figs. \ref{fig4}(b) and \ref{fig4}(c)).
This is the relatively low density regime in which G band phonons
are damped via decay into vertical particle-hole pairs.

\par
The total change in  $\Gamma_G$ is written as $\triangle\Gamma_G =
\Gamma_G(V_{Dirac}) - \Gamma _G ^0$, where $\Gamma _G ^0$ is the
residual line-width from processes that are not related to Landau
damping. The Landau damping phonon decay rate $\triangle\Gamma_G$ is
the probability of creating energetically allowed electron-hole
pairs via the electron-phonon coupling in unit time. Therefore, the
electron-phonon coupling strength can be estimated from
$\triangle\Gamma_G$. A calculation based on the Fermi golden rule
yields \cite{Ferrari2}:
\begin{equation}
\centering
\triangle\Gamma_G=\frac{A_{uc}}{8Mv_F^2}D^2
\end{equation}
where $A_{uc}$ is the area of the graphene unit cell, $\textit{M}$
is carbon atom mass, and $D$ is the electron-phonon coupling
strength. The measured 8.5 cm$^{-1}$ $\triangle\Gamma_G$ yields
$D=14.1~\textrm{eV/\AA}$, which seems to be consistent with the
value predicted by density functional theory \cite{compare}.

\par
The results in Fig. \ref{fig4}(a) exhibit that the measured
$\Gamma_G$ has departures from the step-like function (blue dashed
line) expected for ideal graphene with uniform charge density
distribution. In graphene samples, however, there are presumably
local density variations due to self-doping \cite{Peres}, chemical
adsorbants \cite{Geim1} and charged impurities trapped in the
substrate \cite{Yanwen}. Different locations of the graphene sheet
have thus different local contributions to damping of G phonons,
smoothing out the sudden changes at $\pm\hbar\omega_G/2$.
Considering a non-uniform spatial variation of $\sim\pm~3\times
10^{11}$ cm$^{-2}$ in charge density, we obtained a reasonable fit
for $\Gamma_G$ shown as the red solid curve in Fig. \ref{fig4}(a).

\par
The EFE modulation of $\omega_G$, that has a minimum near the Dirac
point, is shown in Fig. \ref{fig4}(d) as a function of Fermi energy.
For carrier densities above $\sim\pm~5\times10^{11}$ cm$^{-2}$ where
non-uniformity is no longer dominant, the G-band phonon energy
increases linearly with changes in Fermi energy. This intriguing
result directly links the EFE-modulated $\omega_G$ to 2D Dirac
fermion properties, as discussed below.

\par
The coupling between G phonon lattice vibrations and Dirac fermions
is allowed by graphene lattice symmetry. The carriers residing in
the honeycomb lattice respond to the dynamical perturbation of G
mode lattice vibration by creation and annihilation of virtual
long-wavelength electron-hole pairs across the gapless Dirac point,
which in turn, renormalizes the G phonon energy. The energy range of
the virtual electron-hole pairs allowed by Pauli principle is
decided by the position of Fermi level (Figs. \ref{fig4}(e) and
\ref{fig4}(f)). When the graphene is charge neutral, the onset
energy is zero. If graphene is doped with electrons or holes, the
onset energy is twice the Fermi energy. The difference of the G band
energy in neutral and charged graphene is thus given by the
renormalization of electron-hole pairs with energy ranging from 0 to
2$|E_F|$.

\par
Carrying out the calculation explicitly with time dependent
perturbation theory \cite{Allen} and the linear dispersion near the
Dirac point, we find that the change of G-band phonon energy is
described by
\begin{equation}
\centering
\hbar\omega_G-\hbar\omega_G^0=\lambda\{|E_F|+\frac{\hbar\omega_G}{4}\ln|\frac{2|E_F|-\hbar\omega_G}{2|E_F|+\hbar\omega_G}|\}
\end{equation}
where $\omega_G^0$ is $\omega_G$ at Dirac point,
$\lambda=\frac{A_{uc}D^2}{2\pi\hbar\omega_GMv_F^2}$
 \cite{log}. In the high density
limit, the linear term dominates over the logarithm term. A fit of
the linear segments in Fig. \ref{fig4}(d) (blue dashed lines) with
the linear term in Eq. (2) gives $D = 12.6~ \textrm{eV/\AA}$, close
to the value 14.1 $\textrm{eV/\AA}$ determined above from the
analysis of the $\omega_G$ data. The red smooth line in Fig.
\ref{fig4}(d) is the fit that takes into account the impact of
non-uniformity described above.

\par
It is worth noting that
$\frac{\Delta\Gamma_G}{\lambda}=\frac{\pi\hbar\omega_G}{4}\approx$154
meV, which depends only on one parameter. Our experimental result
gives 195 meV.

\par
Finally, we consider the impact of the EFE on the D* Raman mode
shown in Fig. \ref{fig2}(c). Although both peak position and width
of the G and D* bands exhibit similar gate dependence behavior, the
magnitudes of the changes seen in Fig. \ref{fig2}(c) are only
$\sim$10\% of those of the G band. This smaller gate response can be
explained by the impact of electron-phonon interactions for the two
large wavevector phonons that contribute to the second-order Raman
D* band.  The renormalization of the phonon frequency due to
electron-phonon interactions is inversely proportional to the energy
of particle-hole transitions at the phonon wavevector \cite{Allen}.
In this context, the weaker response of the D* band to the EFE
suggests that relevant EFE-modulated particle-hole transitions here
have energies about an order of magnitude larger than those in the
renormalization of the G band \cite{strength}. Further quantitative
description could be carried out by the analysis of the
doubly-resonant Raman processes in the D* band \cite{Thomsen}.

\par
In conclusion, we observed EFE-tunable electron-phonon coupling in
Raman spectra of single layer graphene. The gate voltage dependences
of phonon frequency and damping reveal charge-tunable interactions
of optical phonons with Dirac fermion transitions across a vanishing
bandgap. These results uncover physics that links the EFE with
electron-phonon coupling and demonstrate venues to probe fundamental
interactions and to characterize atomic scale structures and
devices.

\par
\textit{Additional Remark:} During the preparation of this
manuscript, we became aware of related work on similar systems from
other groups \cite{Castro,Ando2,Mauri1,Mauri2}.

\par
We thank I. L. Aleiner, M. S. Hybertsen, A. J. Millis, S. Reich, J.
Lin, J. Maultzsch, Y. Ahmadian and D. Basko for stimulating
discussions. We acknowledge financial support from Nanoscale Science
and Engineering Initiative of the National Science Foundation under
NSF Award Number CHE-0117752 and CHE-0641523, the New York State
Office of Science, Technology, and Academic Research (NYSTAR), and
the Office of Naval Research under Award Number N000140610138. P. K.
acknowledges the support from FENA MARCO Center. A. P. is supported
by the National Science Foundation under Award Number DMR-0352738,
by the Department of Energy under Award Number DE-AIO2-04ER46133,
and by a research grant of the W. M. Keck Foundation.


\begin{thebibliography}{00}
\bibitem{log}The 2nd term of Eq. (2) predicts a logarithmic divergence
as $|E_F|$ approaches $\hbar\omega_G/2$. In the experimental data,
there is no obvious indication of such a divergence, most likely
because it is smoothed out by broadening due to non-uniformity.
\bibitem{strength} The strength of electron-phonon coupling is similar for G
and D* phonons. See Eq. (11) of Ref. \cite{Ferrari1}.
\end{thebibliography}
\end{document}